\documentclass[journal]{IEEEtran}%
\usepackage{amsfonts}
\usepackage{amsmath}
\usepackage{amssymb}
\usepackage{graphicx}%
\providecommand{\U}[1]{\protect\rule{.1in}{.1in}}

\begin{document}
%
\title
{Exact Method for Determining Subsurface Radioactivity Depth Profiles from Gamma Spectroscopy Measurements}%
%

\author{Clinton DeW. Van Siclen
\thanks{
Manuscript received 2011. This work was supported in part by OPW funds from the Idaho National Laboratory, a DOE laboratory operated by Battelle Energy Alliance under DOE Idaho Operations Office Contract DE-AC07-05ID14517.}
\thanks
{C. DeW. Van Siclen is with the Nuclear Nonproliferation Division, Idaho National Laboratory, P. O. Box 1625, Idaho Falls, ID 83415 USA (e-mail: clinton.vansiclen@inl.gov).}%
}%
%

\IEEEpubid{}
\IEEEspecialpapernotice{}%
%

\maketitle
%

\begin{abstract}

Subsurface radioactivity may be due to transport of radionuclides from a
contaminated surface into the solid volume, as occurs for radioactive fallout
deposited on soil, or from fast neutron activation of a solid volume, as
occurs in concrete blocks used for radiation shielding. For purposes including
fate and transport studies of radionuclides in the environment,
decommissioning and decontamination of radiation facilities, and nuclear
forensics, an \textit{in situ}, nondestructive method for ascertaining the
subsurface distribution of radioactivity is desired. The method developed here
obtains a polynomial expression for the radioactivity depth profile, using a
small set of gamma-ray count rates measured by a collimated detector directed
towards the surface at a variety of angles with respect to the surface normal.
To demonstrate its capabilities, this polynomial method is applied to the
simple case where the radioactivity is maximal at the surface and decreases
exponentially with depth below the surface, and to the more difficult case
where the maximal radioactivity is below the surface.%

\end{abstract}
%


\section{Introduction}

Nondestructive methods to determine the extent of subsurface radioactivity
rely on the attenuation of characteristic gamma-rays emitted from the
subsurface. Assuming the radionuclide distribution is reasonably uniform in
the two dimensions parallel to the surface, the problem reduces to finding the
(one-dimensional) depth profile of the radioactivity. To accomplish this, two
general approaches have been taken: (i) A depth distribution function is
\textit{assumed} (from physical considerations), and spectroscopic
measurements are made to determine the best-fit values for the function
parameters. (ii) No depth distribution is assumed; rather, the subsurface
volume is divided into voxels, and spectroscopic measurements are made to
determine the activity of each voxel.

Often there is good reason to assume a particular depth distribution function.
For example, Smith and Elder \cite{r1} consider several parameterized
functions based on physical processes related to radionuclide mobility in
soils. The review by MacDonald \textit{et al.} \cite{r2} discusses three
spectroscopic methods used to obtain the function parameters: (i) The
\textquotedblleft multiple peak method\textquotedblright, which uses the fact
that the scatter cross-section of a gamma-ray is a function of its energy (so
for example, the two parameters of an exponential decay function representing
the radionuclide distribution are calculated from measurements of the count
rates of two characteristic gamma-rays of different energy). (ii) The
\textquotedblleft peak to valley\ method\textquotedblright, which uses the
fact that the ratio of counts in a characteristic gamma-ray peak to counts in
the adjacent \textquotedblleft valley\textquotedblright\ of slightly lower
energy (due to forward scattering of the characteristic gamma-rays) decreases
with increased attenuation of the characteristic gamma-rays (e.g. Gering
\textit{et al.} \cite{r3} who use a two-parameter Lorentz function for
$^{137}$Cs in soil). (iii) The \textquotedblleft lead plate
method\textquotedblright, where an absorbing lead plate inserted between the
detector and the material surface allows the flux contribution from large
angle (with respect to the surface normal) to be distinguished from that from
small angle (e.g., Korun \textit{et al.} \cite{r4} who use a three-parameter
Gaussian function for $^{137}$Cs and $^{134}$Cs in soil; Dewey \textit{et al.}
\cite{r5} who fit a variety of analytical functions to simulated measurements
for $^{137}$Cs in soil; Benke and Kearfott \cite{r6},\cite{r7} who consider
planar radiation sources at various depths).

When a particular depth distribution cannot be assumed, or none of a set of
parameterized functions can be well matched to the spectroscopic measurements,
the subsurface volume may be (virtually) divided into voxels, and the activity
of each voxel determined by some variation of the following procedure. Count
rates of a characteristic gamma-ray are obtained by a collimated detector
directed at (or otherwise sensitive only to gamma-rays emitted at) particular
angles $\theta$ with respect to the surface normal. By obtaining count rates
for a number of different detector orientations, the response matrix equation
$C=\mathbf{R}\Lambda$ can be constructed, where element $c_{i}$ of vector $C$
is the count rate obtained for angle $\theta_{i}$, the detector response
matrix element $r_{ij}$ is the probability that a gamma-ray emitted from voxel
$j$ will contribute to the count rate $c_{i}$ (so $r_{ij}$ accounts for
scattering of that gamma-ray en route to the detector), and element $a_{j}$ of
vector $\Lambda$ is the activity of voxel $j$. Given the vector $C$, the
matrix equation is solved for the activity vector $\Lambda$. This
\textquotedblleft voxel approach\textquotedblright\ is taken by, e.g., Benke
\textit{et al.} \cite{r8},\cite{r9}; Whetstone \textit{et al.} \cite{r10};
Charles \textit{et al.} \cite{r11}.

These approaches to finding the radioactivity depth profile are not entirely
satisfactory for the following reasons. In the case of the \textquotedblleft
parameterized function approach\textquotedblright, the assumption of a
particular depth distribution function is risky in general, even when best-fit
parameter values are found that seem to bear out the assumption. By the nature
of the measurements, the candidate functions are limited to those with very
few parameters, so reasonable best-fit parameter values may be found for more
than one of those functions. In the case of the \textquotedblleft voxel
approach\textquotedblright, the limited number of spectroscopic measurements
greatly limits the number of voxels into which the subsurface is divided. In
any event, the calculated activity $a_{j}$ (gamma-ray emissions per unit time)
is evenly distributed over the entire volume of voxel $j$, so giving the
calculated distribution a spatial resolution no better than the voxel size.

The new approach presented below is motivated by the expectation that the
\textquotedblleft true\textquotedblright\ depth profile function is smooth and
continuous. Thus it can be closely approximated by a polynomial expression,
with more or fewer terms depending on the complexity of the \textquotedblleft
true\textquotedblright\ function (for example, more terms are needed to
approximate a Gaussian-like function than an exponential-like function). The
success of this method is due to the novel way the coefficients of the terms
in the polynomial are obtained from the spectroscopic measurements.

\section{Calculation of the polynomial profile function}

For convenience, the surface is at $z=0$ and the radioactive material lies in
the region $z\geq0$. In what follows, $\rho(z)dV$ is the activity (number of
gamma-rays emitted per second) of the infinitesimal volume $dV$ located in the
subsurface layer $\left[  z,z+dz\right]  $; thus the distribution $\rho(z)$ is
the activity per unit volume at depth $z$.

This \textquotedblleft polynomial method\textquotedblright\ is most easily
developed by considering the idealized case of a highly collimated, one-pixel
gamma-ray detector pointing at the material surface. Due to the collimation,
only those characteristic gamma-rays emitted from a thin \textquotedblleft
pencil\textquotedblright\ of material will be counted at the detector. Unless
all radioactivity is on the surface, the detector counts will provide an
underestimate of the total activity, due to attenuation (scattering) of the
subsurface emissions. Different orientations of the detector with respect to
the surface normal will thus produce different count rates, due to the
different orientations of the pencil\ of material with respect to the
subsurface radioactivity profile. As shown below, that radioactivity profile
can be ascertained from the variation in count rate with detector orientation.

Consider that the highly collimated, one-pixel detector is pointing at the
surface at an angle $\theta$ from the surface normal. The collimation ensures
that all recorded characteristic gamma-rays originate from the thin pencil\ of
material angled by $\theta$ with respect to the surface normal. Note that a
gamma-ray emitted at depth $z$ and recorded at the detector has travelled a
distance $z/\cos\theta$ to the surface. Thus the count rate $d(\theta)$
recorded at the detector is given by%
\begin{equation}
d\left(  \theta\right)  =\varepsilon\int\limits_{z=0}^{z=\infty}\rho
(z)\frac{A}{4\pi\left(  \frac{z+R}{\cos\theta}\right)  ^{2}}\exp\left[
-\mu\left(  \frac{z}{\cos\theta}\right)  \right]  \frac{A}{\cos\theta}dz
\label{e1}%
\end{equation}
where the (unknown) function $\rho(z)$ is the activity per unit volume at
depth $z$; $\mu$ is the attenuation coefficient for the characteristic
gamma-ray in the homogeneous material (attenuation in air is ignored); $R$ is
the normal distance of the pixel center above the surface; $A$ is the
cross-sectional area of the pencil volume, as well as the area of the detector
pixel; and $\varepsilon$ is the detector efficiency (possibly multiplied by
factors related to the system geometry and instrument characteristics). Note
that the integration is over all infinitesimal volumes $dV=dz\cdot
A/\cos\theta$ comprising the pencil volume. The ratio $A\left[  4\pi\left(
\frac{z+R}{\cos\theta}\right)  ^{2}\right]  ^{-1}$ is the fraction of
emissions from the volume $dV$ that would reach the detector in the absence of
attenuation, while the exponential factor accounts for the attenuation.

When count rates are obtained for several different detector orientations
(several different $\theta$ values), a set of these integral equations is
created, which in principle can be solved for the unknown radioactivity
profile $\rho(z)$. The approach taken here is to approximate $\rho(z)$ by the
polynomial $\rho^{\ast}(z)$ of degree $N$,%
\begin{equation}
\rho^{\ast}(z)=\sum\limits_{j=0}^{N}c_{j}\cdot z^{j}=c_{0}+c_{1}z+c_{2}%
z^{2}+\ldots+c_{N}z^{N} \label{e2}%
\end{equation}
where the coefficients $c_{j}$ are determined from the detector count rates in
the following manner.

Each detector orientation with respect to the surface normal produces a linear
equation%
\begin{equation}
d_{i}=\sum\limits_{j=0}^{N}K_{ij}c_{j} \label{e3}%
\end{equation}
where $d_{i}$ is the count rate $d\left(  \theta_{i}\right)  $, as the index
$i $ distinguishes the different detector orientations; and the coefficients%
\begin{equation}
K_{ij}=\varepsilon\int\limits_{z=0}^{z=\zeta}z^{j}\frac{A}{4\pi\left(
\frac{z+R}{\cos\theta_{i}}\right)  ^{2}}\exp\left[  -\mu\left(  \frac{z}%
{\cos\theta_{i}}\right)  \right]  \frac{A}{\cos\theta_{i}}dz\text{.}
\label{e4}%
\end{equation}
Note that the limit of integration $\zeta$ is the depth below which no
radioactivity is present; this is a parameter that must be set (i.e., assumed)
prior to the calculations. In the case of $n$ detector orientations, $n$
equations like Eq. (\ref{e3}) are created. Two additional linear equations are
needed to force the function $\rho^{\ast}(z)$ to go smoothly to zero at
$z=\zeta$. These are%
\begin{equation}
d_{n}=0=\sum\limits_{j=0}^{N}K_{n,j}c_{j} \label{e5}%
\end{equation}
where the coefficients $K_{n,j}=\zeta^{j}$ [this is just the equation
$\rho^{\ast}(\zeta)=0$], and%
\begin{equation}
d_{n+1}=0=\sum\limits_{j=0}^{N}K_{n+1,j}c_{j} \label{e6}%
\end{equation}
where the coefficients $K_{n+1,j}=j\cdot\zeta^{j-1}$ [this is the equation
$d\rho^{\ast}(z)/dz=0$ at $z=\zeta$, which ensures the slope of $\rho^{\ast
}(z)$ is zero at $z=\zeta$].

The set of $n+2$ linear equations is then solved for the set $\left\{
c_{j}\right\}  $, thus producing the radioactivity profile $\rho^{\ast}(z)$
according to Eq. (\ref{e2}). This function reproduces the set $\left\{
d_{i}\right\}  $ of count-rate measurements, and goes smoothly to zero at
$z=\zeta$.

Note that the total activity $\Lambda_{T}$ (number of characteristic
gamma-rays emitted per second) of a \textquotedblleft core\textquotedblright%
\ of cross-sectional area $A$ (taken perpendicular to the surface) is%
\begin{align}
\Lambda_{T}  &  =A\int\limits_{z=0}^{z=\infty}\rho(z)dz\nonumber\\
&  \approx A\int\limits_{z=0}^{z=\zeta}\rho^{\ast}(z)dz=A\sum\limits_{j=0}%
^{N}\frac{c_{j}}{j+1}\zeta^{j+1}\text{.} \label{e7}%
\end{align}
The activity of any slice of that core is obtained by integrating $\rho^{\ast
}(z)$ over the thickness of that slice.

A computational caution: When the integrals in Eq. (\ref{e4}) are evaluated to
obtain the set $\left\{  K_{ij}\right\}  $, it may be found that those values
range over several orders of magnitude. This makes it exceedingly difficult to
solve the system of $n+2$ equations for the set $\left\{  c_{j}\right\}  $ by
conventional methods. Thus an analytical, \textquotedblleft term elimination
sequence\textquotedblright\ method for solving a square system of linear
equations was derived for this purpose. It is presented in the appendix in a
form that is very easy to write into a computer code. All of the polynomial
functions plotted in the following section were obtained by use of this method
incorporated into a simple Mathematica code.

After this exposition of the polynomial method, it is interesting to briefly
consider the \textquotedblleft voxel approach\textquotedblright\ to
determining the radioactivity depth profile mentioned in Section I. The set
$\left\{  d_{i}\right\}  $ of count-rate measurements is related to the set
$\left\{  a_{j}\right\}  $ of voxel activities by a set of linear equations of
the sort%
\begin{equation}
d_{i}=\sum\limits_{j=0}^{N}R_{ij}a_{j}%
\end{equation}
where the $N+1$ voxels are each of size $A\cdot\delta z$ and the dimensionless
coefficients%
\begin{equation}
R_{ij}=\varepsilon\frac{A}{4\pi\left(  \frac{\delta z\cdot j+R}{\cos\theta
_{i}}\right)  ^{2}}\exp\left[  -\mu\left(  \frac{\delta z\cdot j}{\cos
\theta_{i}}\right)  \right]  \frac{1}{\cos\theta_{i}}\text{.}%
\end{equation}
Clearly the number of voxels cannot exceed the number of measurements
(detector orientations), and the voxel thickness $\delta z\gg dz$ must be
sufficient that the array of voxels fills the assumed radioactive volume. Of
course, the polynomial method completely avoids discretization of that volume.

\section{Application of the polynomial method}

Unfortunately, the functional requirements on $\rho^{\ast}(z)$ that it be
continuous, go smoothly to zero at some depth $\zeta$, and reproduce the
spectroscopic measurements $\left\{  d_{i}\right\}  $, do not automatically
ensure that the unknown radioactivity profile function $\rho(z)$ is
reproduced. The finite number of terms in the polynomial (limited by the
number of measurements taken) may be insufficient to produce a complex
function, or the finite depth $\zeta$ at which radioactivity is presumed
negligible may be poorly chosen. Indeed, the exponential decay function, which
is a typical radioactivity profile, is actually a polynomial with an infinite
number of terms that goes to zero at infinity.

\begin{figure}[ptb]%
\centering
\includegraphics[
height=2.6654in,
width=3.1825in
]%
{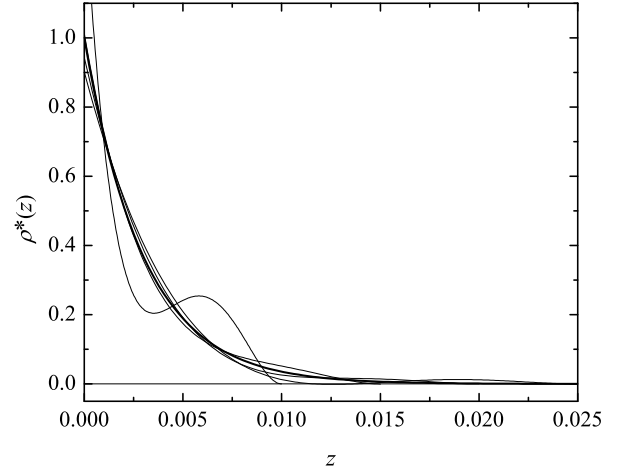}%
\caption{Fourth-degree polynomial representations of the radioactivity depth
profile $\rho\left(  z\right)  =\exp\left[  -z/0.003\right]  $ (the thickest
curve). The odd curve is $\rho^{\ast}\left(  z\right)  $ that terminates at
$\zeta=0.01$; the $\rho^{\ast}\left(  z\right)  $ curves for $\zeta
=0.015,0.02,0.025$ track the $\rho\left(  z\right)  $ curve fairly well. All
the $\rho^{\ast}\left(  z\right)  $ curves are calculated from the count-rate
measurements $\left\{  d\left(  i\frac{\pi}{16}\right)  \right\}  $.}%
\label{fig1}%
\end{figure}

\begin{figure}[ptb]%
\centering
\includegraphics[
height=2.6654in,
width=3.1825in
]%
{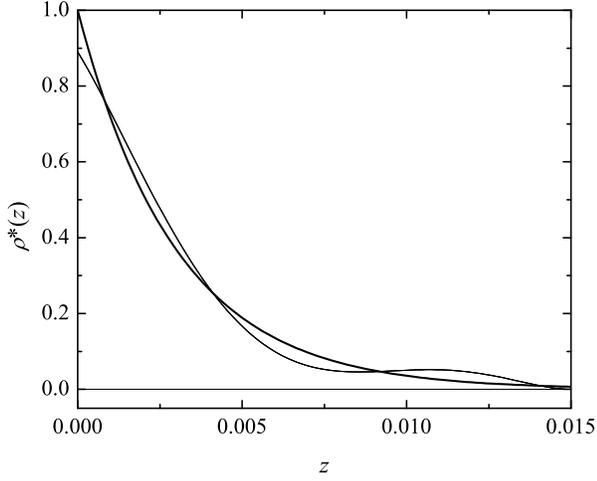}%
\caption{Fifth-degree polynomial representations of the radioactivity depth
profile $\rho\left(  z\right)  =\exp\left[  -z/0.003\right]  $ (the thickest
curve). The $\rho^{\ast}\left(  z\right)  $ curve is actually \textit{three}
indistinguishable curves (for $\zeta=0.015$) calculated from count-rate
measurements $\left\{  d\left(  i\frac{\pi}{20}\right)  \right\}  ,\left\{
d\left(  i\frac{\pi}{16}\right)  \right\}  ,\left\{  d\left(  i\frac{\pi}%
{12}\right)  \right\}  $, respectively.}%
\label{fig2}%
\end{figure}

\begin{figure}[ptb]%
\centering
\includegraphics[
height=2.6654in,
width=3.1825in
]%
{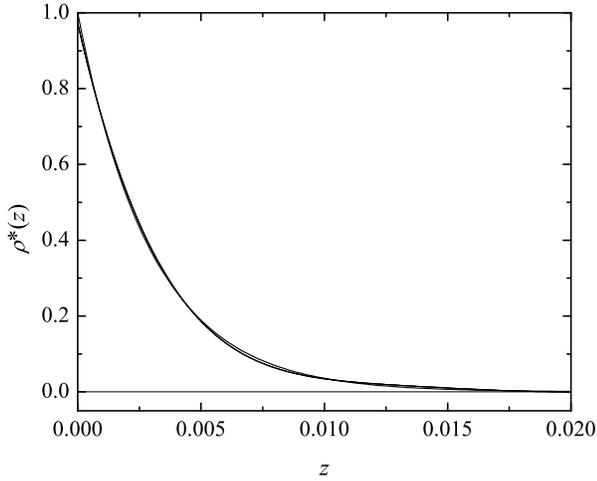}%
\caption{Similar to Fig. \ref{fig2}, but the \textit{three} indistinguishable
$\rho^{\ast}\left(  z\right)  $ curves terminate at $\zeta=0.02$ rather than
at $\zeta=0.015$. The absence of the non-physical hump evident in the curves
of Fig. \ref{fig2} suggests that the 5th degree polynomial $\rho^{\ast}\left(
z\right)  $ for $\zeta=0.02$ is a good representation of the radioactivity
depth profile $\rho\left(  z\right)  $.}%
\label{fig3}%
\end{figure}

But fortunately, non-physical (nonsensical) $\rho^{\ast}(z)$ functions are
easy to distinguish visually. In addition to the functional requirements on
$\rho^{\ast}(z)$, the areas under the curves $\rho^{\ast}(z)$ and $\rho(z)$
are approximately equal [see Eq. (\ref{e7})]. Thus either the curve
$\rho^{\ast}(z)$ closely tracks the curve $\rho(z)$, or the constraints are
accommodated by anomalous undulations of the curve $\rho^{\ast}(z)$. Examples
of this behavior are evident in Fig. \ref{fig1},%
where the radioactivity profile $\rho(z)=\exp[-z/0.003]$ is approximated by a
4th degree polynomial calculated from count-rate measurements at three
detector orientations $\theta=0,\pi/16,\pi/8$. The four curves are $\rho
^{\ast}(z)$ obtained for $\zeta=0.01,0.015,0.02,0.025$, respectively. Clearly
the curve obtained for $\zeta=0.01$ is physically unlikely due to its hump,
while the curve obtained for $\zeta=0.025$ actually goes slightly negative. In
contrast, the curves obtained for $\zeta=0.015$ and $0.02$ look physically
reasonable. Figures \ref{fig2}%
and \ref{fig3}%
show the 5th degree polynomial $\rho^{\ast}(z)$ for $\zeta=0.015$ and $0.02$,
respectively. Each figure shows three curves (impossible to distinguish since
they're essentially identical); each curve is calculated from count-rate
measurements at detector orientations $\theta=0,\varphi,2\varphi,3\varphi$,
where $\varphi$ is $\pi/20,\pi/16,\pi/12$ for the three curves, respectively.
The $\rho^{\ast}(z)$ curves in Fig. \ref{fig2} are physically unlikely (due to
the hump), while there is nothing objectionable about the $\rho^{\ast}(z)$
curves in Fig. \ref{fig3}. Thus any one of the three 5th degree polynomials
(for $\zeta=0.02$) shown in Fig. \ref{fig3} should be a good match to the
unknown profile function $\rho(z)$, as in fact is the case.

Note that the sets $\left\{  d_{i}\right\}  $ [AKA $\left\{  d\left(
\theta_{i}\right)  \right\}  $] of count-rate measurements used to produce the
polynomial curves in these (and following) figures are obtained by solving Eq.
(\ref{e1}) for various values of the detector orientation $\theta$. To specify
those angles, it is convenient to use the notation $\left\{  d\left(
i\frac{\pi}{12}\right)  \right\}  _{i=0,\ldots,N-2}$ (for example) or just
$\left\{  d\left(  i\frac{\pi}{12}\right)  \right\}  $ when the degree $N$ of
the polynomial is otherwise noted (the degree of the polynomial is the number
of detector orientations plus $1$). Also, the calculations use the attenuation
coefficient $\mu=18$ m$^{-1}$ appropriate for the $662$ keV gamma-ray emitted
by $^{137}$Cs in concrete; detector-surface separation $R=0.02$ m; detector
pixel area $A=\left(  0.005\right)  ^{2}$ m$^{2}$; detector efficiency
$\varepsilon=1$.

\begin{figure}[ptb]%
\centering
\includegraphics[
height=2.6654in,
width=3.2059in
]%
{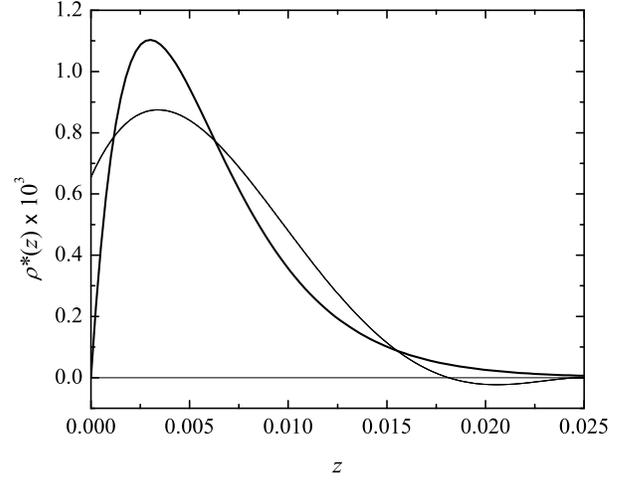}%
\caption{Fourth-degree polynomial representations of the radioactivity depth
profile $\rho\left(  z\right)  =z\exp\left[  -z/0.003\right]  $ (the thickest
curve). The $\rho^{\ast}\left(  z\right)  $ curve is actually \textit{three}
indistinguishable curves (for $\zeta=0.025$) calculated from count-rate
measurements $\left\{  d\left(  i\frac{\pi}{20}\right)  \right\}  ,\left\{
d\left(  i\frac{\pi}{16}\right)  \right\}  ,\left\{  d\left(  i\frac{\pi}%
{12}\right)  \right\}  $, respectively.}%
\label{fig4}%
\end{figure}

\begin{figure}[ptb]%
\centering
\includegraphics[
height=2.6654in,
width=3.2059in
]%
{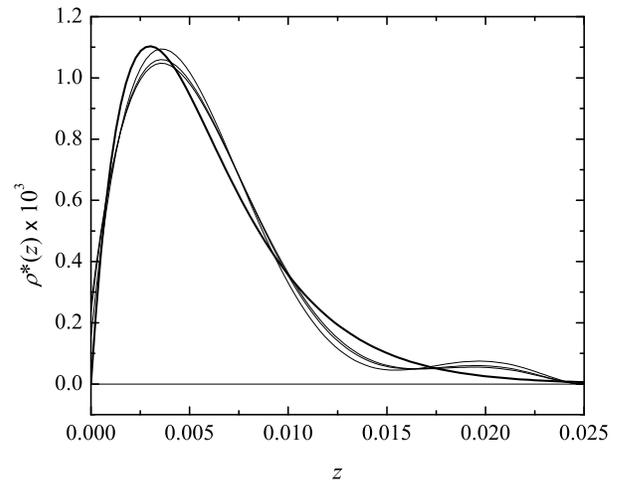}%
\caption{Similar to Fig. \ref{fig4}, but the three $\rho^{\ast}\left(
z\right)  $ curves are \textit{fifth}-degree polynomials.}%
\label{fig5}%
\end{figure}

A more challenging case is that of the radioactivity profile $\rho\left(
z\right)  =z\exp\left[  -z/0.003\right]  $, which goes steeply to zero at the
material surface. A profile of this general shape may occur for neutron
activated concrete shielding, due to thermalization and subsequent capture of
fast neutrons incident on the concrete blocks (Kimura \textit{et al.}
\cite{r12}, Masumoto \textit{et al.} \cite{r13}, Wang \textit{et al.}
\cite{r14}). Figure \ref{fig4}%
shows three (impossible to distinguish) 4th degree polynomials $\rho^{\ast
}(z)$ for $\zeta=0.025$ (other, similar calculations showed the appropriate
value for $\zeta$ to be $0.02<\zeta<0.03$) calculated from count-rate
measurements $\left\{  d\left(  i\frac{\pi}{20}\right)  \right\}  ,\left\{
d\left(  i\frac{\pi}{16}\right)  \right\}  ,\left\{  d\left(  i\frac{\pi}%
{12}\right)  \right\}  $, respectively. These results show convincingly that
the radioactivity is maximal below the surface, but one might wonder about the
behavior very close to the surface, as a 4th degree polynomial has too few
terms to effect a tight bend down to a very low surface value. Figure
\ref{fig5}%
shows the corresponding 5th degree polynomials $\rho^{\ast}(z)$, all of which
go nearly to zero at the material surface.

\begin{figure}[ptb]%
\centering
\includegraphics[
height=2.6654in,
width=3.2059in
]%
{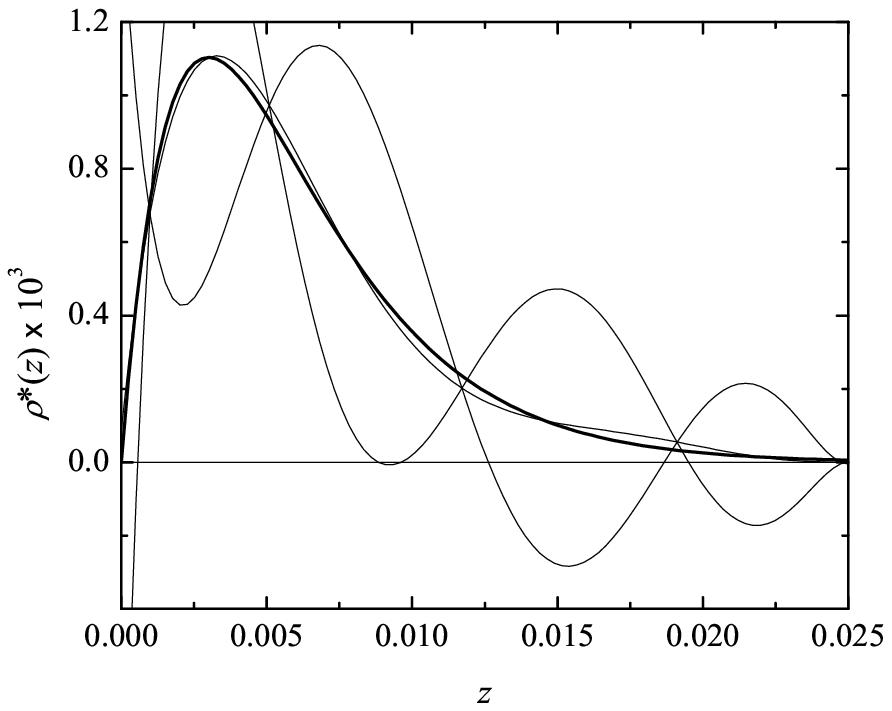}%
\caption{Similar to Figs. \ref{fig4} and \ref{fig5}, but the three $\rho
^{\ast}\left(  z\right)  $ curves are \textit{sixth}-degree polynomials. The
wildly undulating curves calculated from measurements $\left\{  d\left(
i\frac{\pi}{20}\right)  \right\}  $ and $\left\{  d\left(  i\frac{\pi}%
{16}\right)  \right\}  $ are clearly non-physical. In contrast, the reasonable
behavior of the $\rho^{\ast}\left(  z\right)  $ curve calculated from the
measurements $\left\{  d\left(  i\frac{\pi}{12}\right)  \right\}  $ suggests
that it is a good representation of the radioactivity depth profile
$\rho\left(  z\right)  $.}%
\label{fig6}%
\end{figure}

\begin{figure}[ptb]%
\centering
\includegraphics[
height=2.6654in,
width=3.2059in
]%
{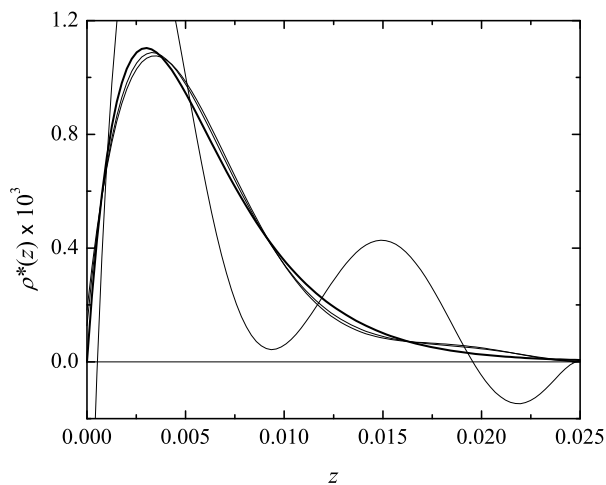}%
\caption{Similar to Figs. \ref{fig4}--\ref{fig6}, but the three $\rho^{\ast
}\left(  z\right)  $ curves are \textit{seventh}-degree polynomials. The
wildly undulating curve is calculated from the measurements $\left\{  d\left(
i\frac{\pi}{20}\right)  \right\}  $ (note that it closely resembles the
corresponding curve in Fig. \ref{fig6}).}%
\label{fig7}%
\end{figure}

\begin{figure}[ptb]%
\centering
\includegraphics[
height=2.6654in,
width=3.2059in
]%
{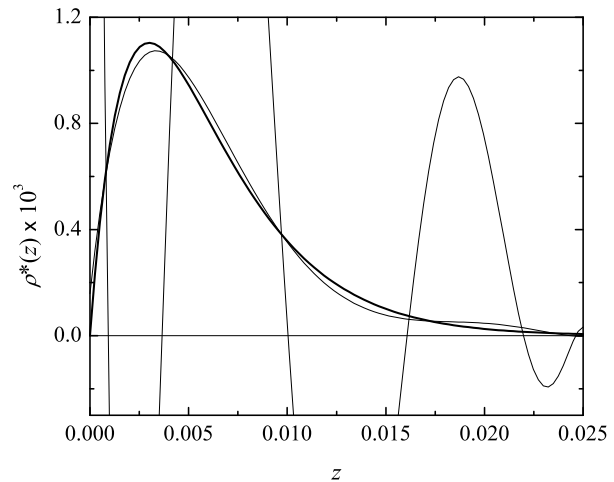}%
\caption{Similar to Figs. \ref{fig4}--\ref{fig7}, but the two $\rho^{\ast
}\left(  z\right)  $ curves are \textit{eighth}-degree polynomials. Again, the
wildly undulating curve is calculated from the measurements $\left\{  d\left(
i\frac{\pi}{20}\right)  \right\}  $, while the reasonable curve is calculated
from the measurements $\left\{  d\left(  i\frac{\pi}{16}\right)  \right\}  $.}%
\label{fig8}%
\end{figure}

It is interesting to consider higher-degree (beyond 5th) polynomial
approximations to the profile function $\rho\left(  z\right)  =z\exp\left[
-z/0.003\right]  $ for their cautionary tales, even though it may be difficult
to get more than four distinct measurements $d_{i}$ in practice. Figure
\ref{fig6}%
shows the corresponding 6th degree polynomials $\rho^{\ast}(z)$. That
calculated from the measurement set $\left\{  d\left(  i\frac{\pi}{12}\right)
\right\}  $ tracks the $\rho\left(  z\right)  $ curve very nicely; the others
are physically impossible! Figures \ref{fig7}%
and \ref{fig8}%
show the corresponding 7th and 8th degree polynomials $\rho^{\ast}(z)$,
respectively. Clearly, use of the set $\left\{  d\left(  i\frac{\pi}%
{12}\right)  \right\}  $ is preferable, presumably due to its wider range of
values (of course, all detector orientations must have $\theta<\pi/2$).

\begin{figure}[ptb]%
\centering
\includegraphics[
height=2.6654in,
width=3.0952in
]%
{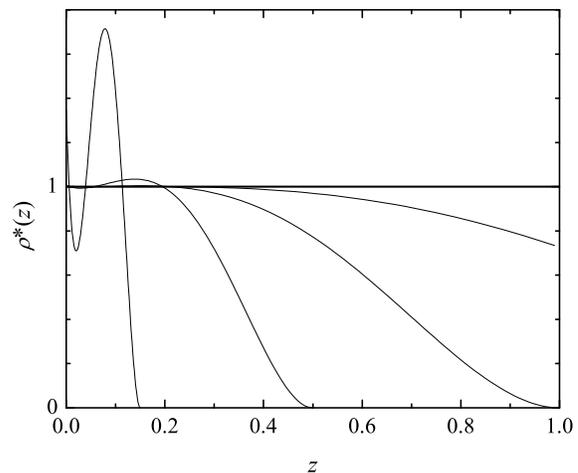}%
\caption{Fourth-degree polynomial representations of the uniform radioactivity
depth profile $\rho\left(  z\right)  =1$ (the thickest curve). The $\rho
^{\ast}\left(  z\right)  $ curves for $\zeta=0.15,0.5,1,2$ are calculated from
the count-rate measurements $\left\{  d\left(  i\frac{\pi}{16}\right)
\right\}  $.}%
\label{fig9}%
\end{figure}

The polynomial method also handles the case of a \textit{uniform}
radioactivity distribution $\rho(z)=1$, as occurs for natural radioactivity in
soils for example. Figure \ref{fig9}%
shows the 4th degree polynomials $\rho^{\ast}(z)$ for $\zeta=0.15,0.5,1,2$
calculated from count-rate measurements $\left\{  d\left(  i\frac{\pi}%
{16}\right)  \right\}  $. From visual inspection of the curves, it is clear
that the polynomial $\rho^{\ast}(z)\rightarrow\rho(z)$ as $\zeta
\rightarrow\infty$.

\section{Use of multi-pixel detectors}

While a \textquotedblleft highly collimated, one-pixel gamma-ray
detector\textquotedblright\ is used in Section II to simplify the presentation
of the method, in practice a collimated detector with many pixels is needed to
accumulate sufficient counts in a reasonable time. Recall that collimation is
necessary so that each pixel records only characteristic gamma-rays emitted at
a particular angle $\theta$ with respect to the surface normal. This is
perhaps most easily accomplished by a collimator that is a thick sheet of lead
with a set of parallel holes (say, $M$ in number) drilled through it. This
effectively converts a standard detector into an array of $M$ collimated,
single-pixel detectors (the single pixel is the size of the hole
cross-section). Of course, there being only a single detector behind the
perforated lead collimator, only a \textit{total} gamma-ray count is acquired;
that is, the \textit{sum} of the contributions of the $M$ single-pixel
detectors is obtained.

Use of such a detector is easily accommodated by the mathematical development
in Section II. In particular, the count rate $d_{i}$ [AKA $d\left(  \theta
_{i}\right)  $, to make clear that the subscript $i$ identifies the particular
value of $\theta$] in Eq. (\ref{e3}) is then the \textit{total} counts
recorded at the detector (that is, the \textit{sum} of the contributions of
the $M$ single-pixel detectors) over the time period, and Eq. (\ref{e4}) for
the coefficients $K_{ij}$ becomes%
\begin{align}
K_{ij}  &  =\varepsilon\int\limits_{z=0}^{z=\zeta}z^{j}\left[  \sum
\limits_{m=1}^{M}\frac{A}{4\pi\left(  \frac{z+R_{i}^{(m)}}{\cos\theta_{i}%
}\right)  ^{2}}\right] \nonumber\\
&  \exp\left[  -\mu\left(  \frac{z}{\cos\theta_{i}}\right)  \right]  \frac
{A}{\cos\theta_{i}}dz \label{e8}%
\end{align}
where $R_{i}^{(m)}$ is the normal distance of pixel $m$ above the surface.

\section{Incorporation of known surface conditions}

The polynomial method is intended for use when the \textquotedblleft
true\textquotedblright\ radioactivity depth profile $\rho\left(  z\right)  $
is unknown. However, it is conceivable that the activity at the material
surface is known (via a swipe test or a shallow core, for example), thus
giving a value for the coefficient $c_{0}$. Then Eq. (\ref{e3}) is written%
\begin{equation}
d_{i}-K_{i0}c_{0}=\sum\limits_{j=1}^{N}K_{ij}c_{j} \label{e9}%
\end{equation}
which is conveniently expressed as%
\begin{equation}
\delta_{i}=\sum\limits_{j=1}^{N}K_{ij}c_{j} \label{e10}%
\end{equation}
where $\delta_{i}\equiv d_{i}-K_{i0}c_{0}$ and the index $i$ runs from $1$ to
$N$, so that the term elimination sequence\ method (see appendix) can be used
to solve the system of equations for the coefficients $c_{1},\ldots,c_{N}$. In
the case of the radioactivity profile $\rho\left(  z\right)  =z\exp\left[
-z/0.003\right]  $ considered above, prior knowledge that $c_{0}=0$ would
allow Eq. (\ref{e10}) to be used, with count-rate measurements $\left\{
d\left(  \left(  i-1\right)  \frac{\pi}{12}\right)  \right\}  _{i=1,\ldots
,N-2}$, say.

Similarly, the slope $c_{1}$ of the radioactivity profile at the surface may
be known as well (for example, buried radionuclides would have $c_{0}=c_{1}=0
$), so Eq. (\ref{e3}) is expressed as%
\begin{equation}
\delta_{i}=\sum\limits_{j=2}^{N}K_{ij}c_{j} \label{e11}%
\end{equation}
where $\delta_{i}\equiv d_{i}-K_{i0}c_{0}-K_{i1}c_{1}$ and the index $i$ runs
from $2$ to $N$, and again the term elimination sequence\ method of solution
can be used.

\section{Concluding remarks}

The expectation that the radioactivity depth profile is continuous and smooth
leads to the \textquotedblleft polynomial method\textquotedblright\ for
determining that profile. Values for the coefficients of the terms in the
polynomial are obtained in a clever way from a set of characteristic gamma-ray
count-rate measurements such that the polynomial function exactly reproduces
the measurements. Practical virtues of this approach include (i) it is
non-destructive, (ii) no \textit{a priori} assumptions regarding the
functional form of the radioactivity depth profile are needed, and (iii) very
few spectroscopic measurements are needed. In contrast, current
non-destructive approaches \textit{assume} a particular radioactivity depth
profile function (which is then fitted to the measurements) or divide the
subsurface volume into a small number of voxels (which cannot exceed the
number of distinct measurements).

The polynomial method is particularly well suited to ascertaining the depth
distribution of subsurface radioactivity due to neutron activation of
radiation shielding (Kimura \textit{et al.} \cite{r12}, Masumoto \textit{et
al.} \cite{r13}, Wang \textit{et al.} \cite{r14}, \v{Z}agar and Ravnik
\cite{r15}) and structural materials (Nakanishi \textit{et al.} \cite{r16}).
Thus important applications are to decommissioning and decontamination of
accelerators and other radiation facilities (e.g., enabling separation of
radioactive portions from larger volumes), and nuclear forensics (e.g.,
determining the neutron fluence and energy spectrum after a nuclear event).

\section*{Appendix: Term elimination sequence method for solution of systems
of linear equations}

After obtaining the set $\left\{  d_{i}\right\}  $ of count-rate measurements
and calculating the set $\left\{  K_{ij}\right\}  $ of coefficients, the
system of equations of the sort Eq. (\ref{e3}) must be solved for the set
$\left\{  c_{j}\right\}  $. Note that%
\[
d_{i}=\sum\limits_{j=0}^{N}K_{ij}c_{j}=K_{iN}c_{N}+\sum\limits_{j=0}%
^{N-1}K_{ij}c_{j}%
\]%
\[
\Rightarrow\frac{d_{N}}{K_{NN}}-\frac{d_{i}}{K_{iN}}=\sum\limits_{j=0}%
^{N-1}\left(  \frac{K_{Nj}}{K_{NN}}-\frac{K_{ij}}{K_{iN}}\right)  c_{j}%
\]%
\begin{equation}
\text{with }i,j\leq N-1\text{.}%
\end{equation}
This last equation can be written%
\[
q_{i}=\sum\limits_{j=0}^{N-1}Q_{ij}c_{j}\text{ where }q_{i}\equiv\frac{d_{N}%
}{K_{NN}}-\frac{d_{i}}{K_{iN}}\text{ and}%
\]%
\begin{equation}
Q_{ij}\equiv\frac{K_{Nj}}{K_{NN}}-\frac{K_{ij}}{K_{iN}}\text{ with }i,j\leq
N-1\text{.}%
\end{equation}
Clearly this two-step procedure can be repeated until the number of terms in
the summation is reduced to one. This suggests the following method of solution.

Consider the (square) system of linear equations of the sort%
\begin{equation}
q_{i}^{(N)}=\sum\limits_{j=0}^{N}Q_{ij}^{(N)}c_{j}%
\end{equation}
where the superscripts on the quantities $q_{i}$ and $Q_{ij}$ are intended
only to distinguish the sets $\left\{  q_{i}^{(N)},Q_{ij}^{(N)}\right\}  $,
$\left\{  q_{i}^{(N-1)},Q_{ij}^{(N-1)}\right\}  $, etc., as needed below. The
quantities $q_{i}^{(N)}$ and $Q_{ij}^{(N)}$ correspond to $d_{i}$ and $K_{ij}
$, respectively, so the set $\left\{  q_{i}^{(N)},Q_{ij}^{(N)}\right\}
_{0\leq i,j\leq N}$ is already known. Then according to the two-step procedure
above,%
\[
q_{i}^{(N-1)}\equiv\frac{q_{N}^{(N)}}{Q_{NN}^{(N)}}-\frac{q_{i}^{(N)}}%
{Q_{iN}^{(N)}}\text{ and}%
\]%
\begin{equation}
Q_{ij}^{(N-1)}\equiv\frac{Q_{Nj}^{(N)}}{Q_{NN}^{(N)}}-\frac{Q_{ij}^{(N)}%
}{Q_{iN}^{(N)}}\text{ with }i,j\leq N-1
\end{equation}%
\[
q_{i}^{(N-2)}\equiv\frac{q_{N-1}^{(N-1)}}{Q_{N-1,N-1}^{(N-1)}}-\frac
{q_{i}^{(N-1)}}{Q_{i,N-1}^{(N-1)}}\text{ and}%
\]%
\begin{equation}
Q_{ij}^{(N-2)}\equiv\frac{Q_{N-1,j}^{(N-1)}}{Q_{N-1,N-1}^{(N-1)}}-\frac
{Q_{ij}^{(N-1)}}{Q_{i,N-1}^{(N-1)}}\text{ with }i,j\leq N-2
\end{equation}%
\[
\vdots
\]%
\[
q_{i}^{(1)}\equiv\frac{q_{2}^{(2)}}{Q_{22}^{(2)}}-\frac{q_{i}^{(2)}}%
{Q_{i2}^{(2)}}\text{ and}%
\]%
\begin{equation}
Q_{ij}^{(1)}\equiv\frac{Q_{2j}^{(2)}}{Q_{22}^{(2)}}-\frac{Q_{ij}^{(2)}}%
{Q_{i2}^{(2)}}\text{ with }i,j\leq1
\end{equation}%
\[
q_{i}^{(0)}\equiv\frac{q_{1}^{(1)}}{Q_{11}^{(1)}}-\frac{q_{i}^{(1)}}%
{Q_{i1}^{(1)}}\text{ and}%
\]%
\begin{equation}
Q_{ij}^{(0)}\equiv\frac{Q_{1j}^{(1)}}{Q_{11}^{(1)}}-\frac{Q_{ij}^{(1)}}%
{Q_{i1}^{(1)}}\text{ with }i,j=0
\end{equation}
In this way all quantities $\left\{  q_{i}^{(N-1)},Q_{ij}^{(N-1)}\right\}
_{0\leq i,j\leq N-1},\ldots,\left\{  q_{i}^{(0)},Q_{ij}^{(0)}\right\}
_{i,j=0} $ are calculated. Then $c_{0},c_{1},\ldots,c_{N}$, respectively, are
obtained from%
\begin{equation}
q_{0}^{(0)}=Q_{00}^{(0)}c_{0}%
\end{equation}%
\begin{equation}
q_{1}^{(1)}=Q_{11}^{(1)}c_{1}+Q_{10}^{(1)}c_{0}%
\end{equation}%
\begin{equation}
q_{2}^{(2)}=Q_{22}^{(2)}c_{2}+\sum\limits_{j=0}^{1}Q_{2j}^{(2)}c_{j}%
\end{equation}%
\[
\vdots
\]%
\begin{equation}
q_{N}^{(N)}=Q_{NN}^{(N)}c_{N}+\sum\limits_{j=0}^{N-1}Q_{Nj}^{(N)}c_{j}%
\end{equation}

\end{document}